\documentclass[]{aa}

\newcommand{\mnras}[1]{MNRAS}
\newcommand{\apj}[1]{ApJ}
\newcommand{\apjl}[1]{ApJL}
\newcommand{\nat}[1]{Nature}
\newcommand{\aap}[1]{A\&A}
\newcommand{\aj}[1]{AJ}

\usepackage{epsf}

\begin{document}

\thesaurus{02(12.04.3; 12.07.1; 11.08.1; 11.17.4 B1600+434; 11.19.6)}

\title{A time-delay determination from VLA light curves of the CLASS
       gravitational lens B1600+434}

\author{L.V.E. Koopmans\inst{1}, A.G. de Bruyn\inst{2,1}, 
        E. Xanthopoulos\inst{3}, C.D. Fassnacht\inst{4}}

\institute{Kapteyn Astronomical Institute, P.O. Box 800, 
           NL--9700 AV Groningen, The Netherlands
	   \and
	   NFRA, P.O. Box 2, 
	   NL-7990 AA Dwingeloo, The Netherlands
	   \and
	   University of Manchester, NRAL Jodrell Bank, Macclesfield, 
	   Cheshire SK11 9DL, UK
	   \and
	   National Radio Astronomy Observatory, PO Box 0, Socorro, NM 87801, USA}

\mail{leon@astro.rug.nl}
\offprints{L.V.E. Koopmans}

\date{\today/\today}

\authorrunning{L.V.E. Koopmans et al.}
\titlerunning{A time-delay from CLASS B1600+434}

\maketitle

\begin{abstract}

We present {\sl Very Large Array} (VLA) 8.5-GHz light curves of the
two lens images of the {\sl Cosmic Lens All Sky Survey} (CLASS)
gravitational lens B1600+434.  We find a nearly linear decrease of
18--19\% in the flux densities of both lens images over a period of
eight months (February-October) in 1998. Additionally, the brightest
image A shows modulations up to $11\%$ peak-to-peak on scales of
days to weeks over a large part of the observing period. Image B
varies significantly less on this time scale.  We conclude that most
of the short-term variability in image A is {\sl not} intrinsic
source variability, but is most likely caused by microlensing in the
lens galaxy.  The alternative, scintillation by the ionized Galactic
ISM, is shown to be implausible based on its strong opposite frequency
dependent behavior compared with results from multi-frequency WSRT
monitoring observations (Koopmans \& de Bruyn 1999).

From these VLA light curves we determine a median time delay between
the lens images of $47^{+5}_{-6}$\,d (68\%) or $47^{+12}_{-9}$\,d
(95\%). We use two different methods to derive the time delay; both
give the same result within the errors.  We estimate an additional
systematic error between $-$8 and +7\,d. If the mass distribution of
lens galaxy can be described by an isothermal model (Koopmans, de
Bruyn \& Jackson 1998), this time delay would give a value for the
Hubble parameter, H$_0$=$57^{+14}_{-11}$ (95\% statistical)
$^{+26}_{-15}$ (systematic)~km s$^{-1}$ Mpc$^{-1}$ ($\Omega_{\rm m}$=1
and $\Omega_{\Lambda}$=0). Similarly, the Modified-Hubble-Profile
mass model would give H$_0$=$74^{+18}_{-15}$ (95\% statistical)
$^{+22}_{-22}$ (systematic)~km s$^{-1}$ Mpc$^{-1}$. For $\Omega_{\rm
m}$=0.3 and $\Omega_{\Lambda}$=0.7, these values increase by 5.4\%. We
emphasize that the slope of the radial mass profile of the
lens-galaxy dark-matter halo in B1600+434 is extremely
ill-constrained.  Hence, an accurate determination of H$_0$ from this
system is very difficult, if no additional constraints on the mass
model are obtained. These values of H$_0$ should therefore be regarded as
indicative.

Once H$_0$ (from independent methods) and
the time delay have been determined with sufficient accuracy, it will
prove more worthwhile to constrain the radial mass profile of the
dark-matter halo around the edge-on spiral lens galaxy at $z$=0.4.


\end{abstract}

\section{Introduction}

Gravitational lens systems can be used to determine the Hubble
parameter, H$_0$ (Refsdal 1964). However, both a good mass model of
the lens galaxy as well as a time delay between an image pair are
necessary ingredients to accomplish this. The mass model can be
constrained from the lens-image properties, whereas a time delay can
be obtained through correlations between two image light curves.
Intrinsic source variability should occur in all light curves, lagging
by time delays which depend on the source and lens redshifts, 
the lens-mass distribution and the cosmological parameters,
most prominently H$_0$ (e.g. Schneider et al. 1992).

Recently, time delays and values of H$_0$ were determined from several
different gravitational lens (GL) systems: Q0957+561 (e.g. Schild \& 
Thomson 1997; Kundi\'c et al. 1997a; Haarsma et
al. 1997, 1999; Press, Rybicki \& Hewitt 1992; Pelt et al. 1994; Pelt
et al. 1996; Grogin \& Narayan 1996; Bernstein et al. 1997; Fischer et
al.  1997; Falco et al. 1997; Bernstein \& Fischer 1999; Barkana et
al. 1999), PG1115+080 (e.g. Schechter et al. 1997; Courbin et
al. 1997; Kundi\'c et al. 1997b; Keeton \& Kochanek 1997; Barkana
1997; Saha \& Williams 1997; Pelt et al. 1998; Impey et al. 1998;
Romanowsky \& Kochanek 1999), B0218+357 (e.g. Biggs et al. 1999),
B1608+656 (e.g. Fassnacht et al. 1999; Koopmans \& Fassnacht 1999) and
PKS~1830-211 (e.g. van Ommen et al. 1995; Lovell et al. 1998; Kochanek
\& Narayan 1992; Nair, Narasimha \& Rao 1993). 

Assuming that the lens galaxies in these GL systems can be described
by an isothermal mass distribution, one finds that the values of H$_0$
derived from these GL systems -- except for PG1115+080 -- are consistent
within their 1-$\sigma$ errors {\sl and} agree with the local, SNe Ia
and S-Z determinations of H$_0$ (e.g. Koopmans \& Fassnacht
1999). However, one has to keep in mind that a similar mass-sheet or
change in the radial mass profile of the lens galaxies introduce a
similar change in the determination of H$_0$ from each of these GL
systems (e.g. Falco, Gorenstein \& Shapiro 1985; Gorenstein, Shapiro
\& Falco 1988; Grogin \& Narayan 1992; Keeton \& Kochanek 1997;
Wucknitz \& Refsdal 1999). Hence, also for mass models other than
isothermal one might expect the values of H$_0$ to agree to first
order. Strong deviations of the lens-galaxy mass distribution from
isothermal, however, would lead to systematic differences between the
values of H$_0$ determined from lensing and those determined from
other methods.

In this paper, we present a determination of the time delay between
the two lens images of the CLASS gravitational lens B1600+434, using
the light curves obtained during an eight-month VLA 8.5-GHz monitoring
campaign.  In section 2, we describe the VLA observations of
B1600+434 and the data reduction. In section 3, we apply the minimum
dispersion method from Pelt et al. (1996) and the PRH-method (Press
et al. 1992) to determine a time delay between the lens images and use that
to estimate a tentative value for H$_0$, keeping the above-mentioned problems
with the mass-model degeneracies in mind. In section 4, our
conclusions are summarized.
 
\section{Data \& Reduction}

\subsection{Observations}

We observed the CLASS gravitational lens B1600+434 (Jackson et
al. 1995; Jaunsen \& Hjorth 1997; Koopmans, de Bruyn \& Jackson 1998)
with the VLA in A- and B-arrays at 8.5 GHz (X-band), during the
period 1998 February 13 to October 14. The typical angular resolution
of the radio images ranges from about 0.2 arcsec in A-array to 0.7
arcsec in B-array, sufficient to resolve the 1.4-arcsec double.  In
total we obtained 75 epochs of about 30 min each, including phase-
and flux-calibrator observations and slewing time. The average time
interval between epochs was 3.3 days. A typical observing run
consisted of the sequence listed in Table~1.  This sequence was
repeated, typically twice, until about 30 min observing time was
filled. Several epochs consisted of only 10 or 20 min, making it
necessary to reduce the number of sequences, or the time spent on each
of the sources. At the end of a sequence, we again observed 2 min on
the phase calibrator J1549+506 (Fig.1; Patnaik et al. 1992).

The flux calibrator, B1634+627 (3C343), is a slightly extended
steep-spectrum source (Fig.1; e.g. van Breugel et al. 1992), which
should not be variable.  Its flux density at 8.5 GHz is 0.84$\pm$0.01\,Jy, 
which we obtained with the {\sl
Westerbork-Synthesis-Radio-Telescope} (WSRT) in December 1998.
This value agrees with the average flux density of 0.83\,Jy, obtained
with the 26-m University of Michigan Radio Astronomy Observatory
(UMRAO) telescope, which has been monitoring this source for the past
15 years. We use the WSRT flux-density determination to bring all
flux densities to the correct absolute scale (Sect.2.4).

\begin{table}[t!]
\caption{Basic observing sequence on B1600+434 and the phase and flux calibrators.
This sequence was repeated typically twice per session.}
\centering
\begin{tabular}{llll}
\hline
1 & J1549+506 & 2 min & phase calibrator\\
2 & B1634+627 & 2 min & flux calibrator\\
3 & B1600+434 & 6 min & CLASS gravitational lens\\
4 & B1634+627 & 2 min & flux calibrator\\
\hline
\end{tabular}
\end{table}

\subsection{Calibration}

The initial flux and phase calibration is done in the NRAO
data-reduction package {\sf AIPS} (version 15OCT98). We fixed the
flux density of the phase calibrator at 1.17\,Jy\footnote{Taken from
the VLA Calibrator Manual.} at all epochs and subsequently solve for
the telescope phase and gain solutions. The phase calibrator
(J1549+506; Patnaik et al. 1992) is an unresolved flat-spectrum point
source ($<$10~mas)\footnote{ See the VLBA Calibrators List.} at 8.5
GHz when observed with the VLA in A-array (Fig.1). A single delta
function is therefore sufficient to describe the source
structure. Because the phase calibrator is a flat spectrum and compact
source, it could vary significantly over a monitoring period of eight
months.  Thus, the flux density of 1.17\,Jy could be wrong by a large
factor.  We therefore use the flux calibrator to determine the
proper flux-density scale and correct for `variability' introduced in
both images of B1600+434, due to intrinsic flux-density variability
of the phase-calibrator. As we will see in Sect.2.4, this strategy works well.

We smooth (1-min intervals) and interpolate the phase and gain
solutions between the phase calibrator scans for the entire
observation period of about 30 min. These are then used to determine
initial phase and gain solutions for the flux calibrator B1634+627 and
the gravitational lens system B1600+434. No flagging was done in {\sf
AIPS}, because it is hard to assess if phase and/or gain errors can be
corrected through self-calibration. Flagging is only done in the {\sf
DIFMAP} package (Shepherd 1997), where the visibility amplitudes and
phases can be compared with a model of the source structure and
possibly be corrected through self-calibration.  Only if the latter
fails, do we manually flag those visibilities which most strongly
deviate from the model.

\begin{figure*}
\begin{center}
  \leavevmode
\vbox{%
  \epsfxsize=\hsize
  \epsffile{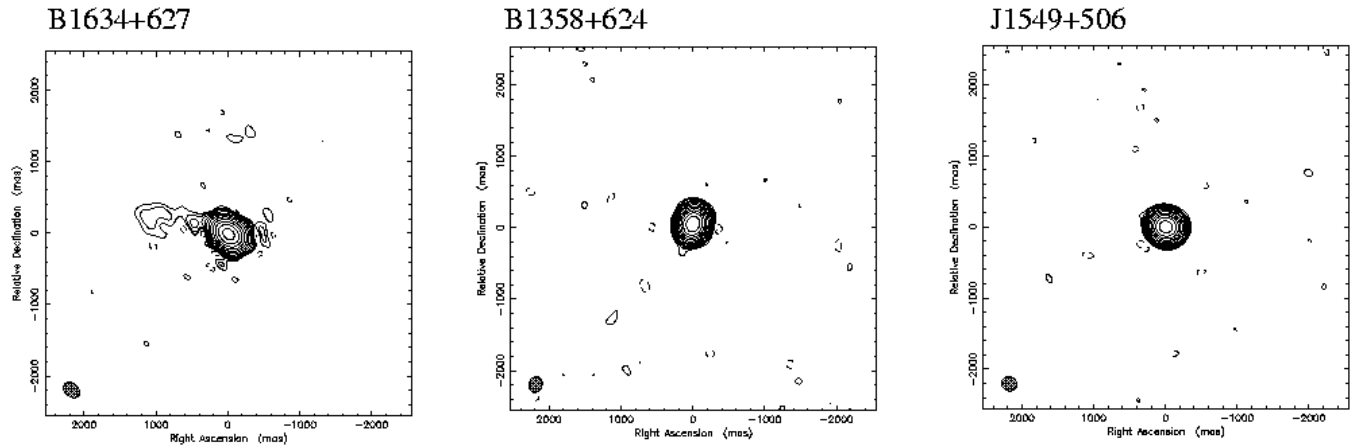}}
\end{center}
\caption{VLA 8.5-GHz A-array images of the flux density calibrators B1634+627
	and B1358+624, and the phase calibrator J1549+506 (epoch 1998 May
	25). Contours starts at 3-$\sigma$ and increase by factors of two.
	The lowest contours are 0.53, 0.56 and 0.56 mJy~beam$^{-1}$ for B1634+627,
	B1358+624 and J1549+506, respectively.}
\end{figure*}

\begin{figure*}[t!]
\begin{center}
  \leavevmode
\vbox{%
  \epsfxsize=\hsize
  \epsffile{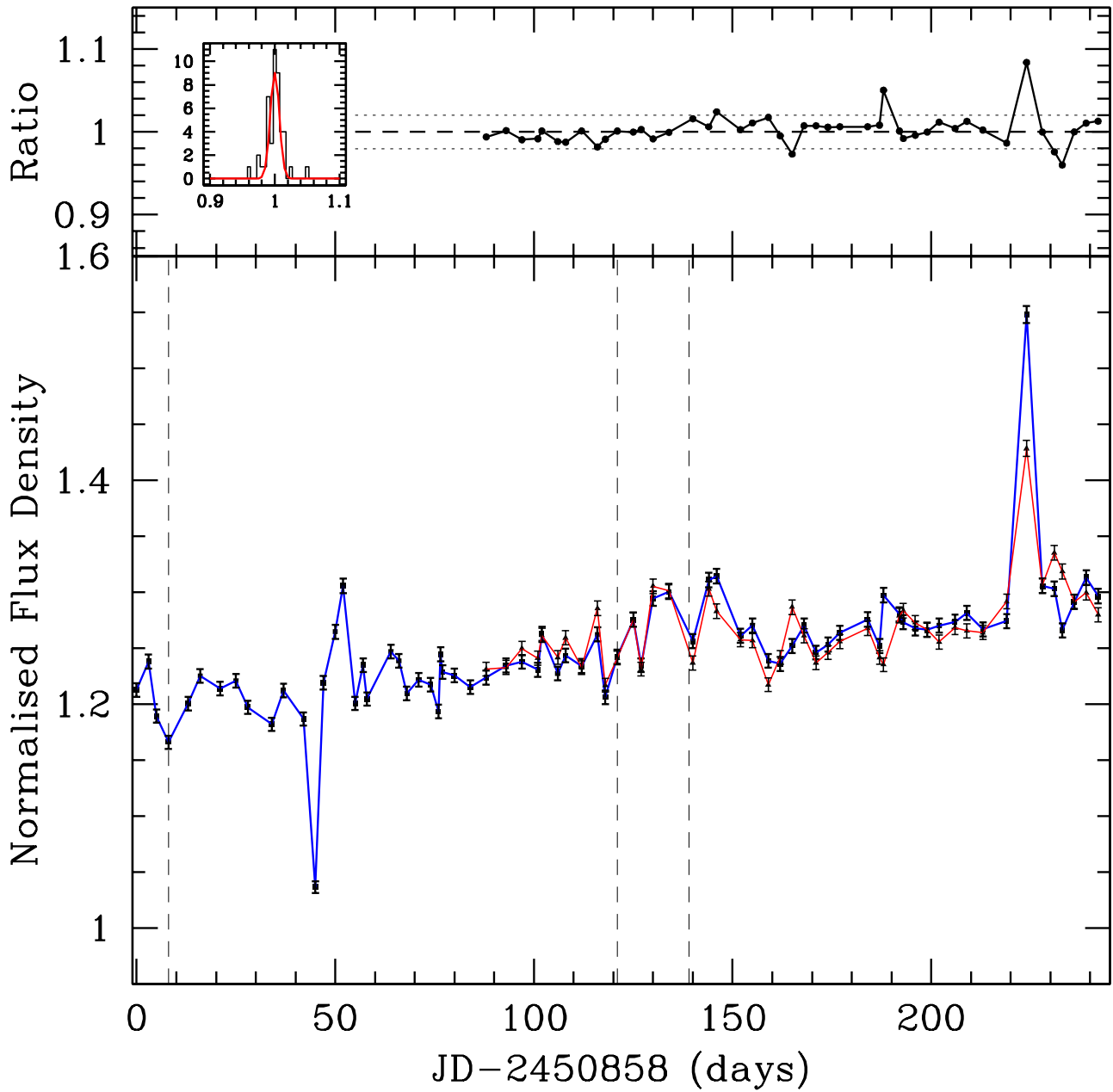}}
\end{center}
\caption{The lower panel shows the light curves of B1634+627 (squares)
	and B1358+624 (triangles), normalized by the flux density of 
        J1549+506 (i.e. 1.17\,Jy; see Sect.2.1).
	The upper panel shows the ratio of the two curves, where they overlap. The
	small sub panel shows a histogram of this ratio, fitted by a Gaussian
	with a 1-$\sigma$ value of $\approx$0.7\%.  The dotted lines in the
	upper panel indicate the $\pm$3-$\sigma$ region from the Gaussian
	fit. The error bars on the light curves are
	0.7\%$/\sqrt{2}\approx0.5\%$ (Sect.2.7), assuming the errors on both
	light curves are similar.  The vertical dashed lines indicate the
	array changes D$\rightarrow$A, A$\rightarrow$BnA and
	BnA$\rightarrow$B, respectively, from left to right. }
\end{figure*}
 
\subsection{Imaging and model fitting}

First we make maps of B1600+434 for all epochs in {\sf DIFMAP}
(version 2.2c), using a process of iterative model fitting and
self-calibration. Because the two lens images are compact ($\le$1
mas) at 8.5-GHz, determined from VLBA observations (Fig.3), and also
show no extended emission from either the lens or quasar on mas to
arcsec scales in MERLIN 5-GHZ observations (Koopmans et al. 1998), we
can safely model the lens image structure by two delta functions, for
which we determine the positions and flux densities. We fit these
delta functions to the visibilities, until the model-$\chi^2$
converges. We perform a phase-only self-calibration, using this
model. Typically, this decreases the $\chi^2$ of the model and the rms
noise in the map significantly. We repeat this process several times,
until no further decrease in either $\chi^2$ or rms noise level is
obtained. Finally, a global gain-self-calibration is performed,
solving for small gain errors of each telescope. This
gain-self-calibration does not significantly change the flux
densities of the images (i.e. changes that are much less than the
statistical error on the flux densities) and individual telescope gain
corrections are typically less than a few percent. We repeat the
process of model fitting and self-calibration, until it converges
again. The visibilities are then compared with the best model and
obviously errant points are manually flagged. Once more, we iterate
between model fitting and self-calibration until convergence is
reached.  Finally, we average the visibilities over 300 sec and repeat
the convergence process, after having flagged those points which still
deviate significantly ($\ga$3-sigma) from the model.

The flux-density ratio between images A and B changes by 
at most a few tenths of a per cent before and after these
calibration cycles, as long as there are no visibilities which deviate
by orders of magnitude from the model.  The reduced-$\chi^2$-value
of the final model fit typically lies between 1.0 and 1.1. The
residual maps show no spurious features due to bad visibilities. Fig.3
shows a radio map of B1600+434, created from the combined A-array data-sets of
seven epochs.

\subsection{Flux calibration}

We subsequently make maps of the flux calibrator, B1634+ 627, using
the same procedure as for B1600+434. 
B1634+627 can be well represented by a single Gaussian component with
a 1-$\sigma$ major axis of 90\,mas, an axial ratio of 0.8 and a
position angle of 70\,deg, in broad agreement with the source
structure as seen in 50-cm VLBI images (Nan et al. 1991). We use
this component to model fit the image structure and self-calibrate the
visibilities. We subsequently remove this Gaussian component and clean
the map to find a better description of the extended structure of the
source. The ratio between the flux density in the Gaussian component
and the total flux density in all extended emission (seen with the VLA 
in A and B-arrays on a scale $\la$10 arcsec) is 0.98. The ratio
is independent of epoch (within the errors). The determination of the
flux density using a single Gaussian component is better defined than
the flux density derived from an iterative cleaning procedure.
The latter procedure seems to introduce $\la$1\% errors, depending
on how the maps were cleaned, what array was used and inside which
box the total flux density was determined. Using only a single Gaussian
to represent the source, does not involve cleaning or choosing a box
size.

We determine the normalized light curve of the flux calibrator
B1634+627 by dividing its flux-density light curve by the flux
density of 0.84$\pm$0.01\,Jy (Sect.2.1). The resulting curve shows a
linear increase of approximately 5\% over the eight-month observing
period, which we attribute to a similar change in the flux density of
the flat-spectrum phase calibrator, J1549+506. We also see that we
have overestimated the flux density of B1634+627 by some 20--30\%. In
other words, our initial estimate of the flux density of the phase
calibrator, based on the value given in the VLA calibrator manual, was
20--30\% too high, which comes as no surprise for a variable flat-spectrum 
radio source.

From May 12 1998 onwards, we added a second flux
calibrator\footnote{Taken from the VLA Calibrator Manual.}, B1358+624
(Fig.1), to the observations, to estimate the reliability of the flux-density 
determination of B1634+627. We followed the same calibration
and mapping procedure as for B1634+627. B1358+624 also shows the same
5\% linear increase in flux density, supporting the idea that this is
the result of variability of the phase calibrator. We scale the flux
density curve of B1358+624 to fit the calibrated light curve of
B1634+627 and find its flux density to be 1.14$\pm$0.02\,Jy.  The
final normalized flux density curve of B1358+624 is also shown in
Fig.2.

\begin{figure}[t!]
\begin{center}
  \leavevmode
\vbox{%
  \epsfxsize=\hsize
  \epsffile{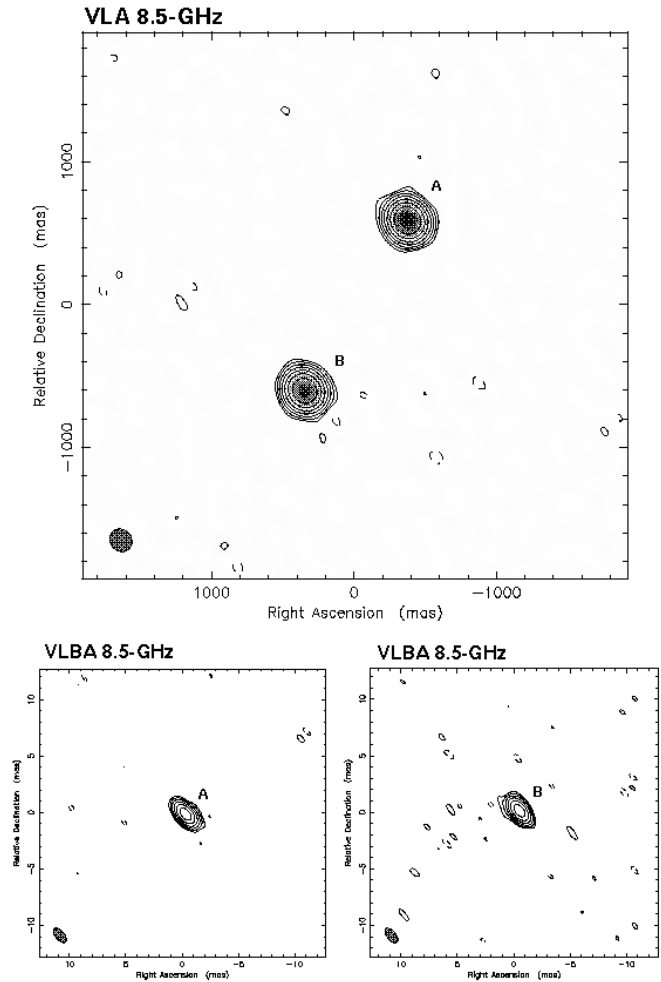}}
\end{center}
\caption{Top: VLA 8.5-GHz A-array image of B1600+434, showing the two
	compact lens images A and B. The image was created from a
	combined data set from 7 different epochs. The contours
	indicate (-3,3,6,12,24,48,96,192,384)$\times$rms noise, where
	the rms noise is 0.059 mJy~beam$^{-1}$. Bottom: VLBA 8.5-GHz images of the
	lens components B1600+434 A and B (5-h integration) taken 1996
	November 4. For a full discussion on the VLBA data
	reduction see Koopmans et al. (1999), where this has been
	explained in detail for B1127+385, which was observed during
	the same run as B1600+434. The contours indicate
	(-2,2,4,8,16,32,64) per cent of the map peak, where the map
	peak is 30.1 (23.4) mJy~beam$^{-1}$ for image A (B).  Both lens images
	show no sign of any extended structure larger than about 1\,mas.}
\end{figure}

\subsection{Light curves of B1600+434}

To correct the light curves of B1600+434 for flux-density calibration
errors, we divide them by the average of the normalized light curves of 
B1634+627 and B1358+624.

We assume that both flux calibrators do not vary intrinsically over the
eight-month observing period. We also assume that the phase and gain
solutions found from J1549+ 506 do not change significantly over a
time span of several minutes, such that interpolation can be used to
make a first-order correction for phase and gain errors in data of
B1600+434, B1634+627 and B1358+624. This flux density correction
removes the largest errors in the light curves of B1600+434, after
which only statistical errors, second-order systematic errors and
intrinsic variations are left. Any deviations between the two
normalized light curves are most likely due to short-term calibration
errors, either instrumental or atmospheric. The final
flux-calibrated light curves of the lens images A and B are shown in
Fig.4. The calibrated flux-densities are listed in Table~2.

Both light curves show a gradual decrease of about 0.02~mJy\,day$^{-1}$
over a period of 243~days (Sect.3.1). Superposed on this gradual
long-term variability, image A also exhibits strong (up to 11\%
peak-to-peak) modulations on a time scale of a few days to several
weeks. Image B only shows a few (up to 6\% peak-to-peak) features, but
separated in time by about one month or more. The modulation indices
(i.e. fractional rms variability) around the gradual long-term
decrease in flux density (indicated by the dashed lines in Fig.4) are
2.8\% and 1.9\% for images A and B, respectively.

\begin{table*}
\caption{The calibrated flux-densities of B1600+434-A and B at
8.5-GHz, during the VLA A and B-arrays monitoring campaign in 1998.
The 1-$\sigma$ errors on the flux densities are 0.8\% for days 0--84
and 0.7\% for days 88--243. The six `outliers' (Sect.2.7) are
indicated by a star (*) and their estimated error ($\sigma_{\rm S}$)
is given in the comments.}
\begin{center}
\begin{tabular}{crccccl}
\hline
  Epoch        & Date & Start    & Array  & $S_{8.5}^{\rm A}$ &  $S_{8.5}^{\rm B}$      & Comments\\
JD$-$2450858     && (LST)    &        &   (mJy)     &    (mJy)         &          \\ 
\hline
    0 &  13/2/98 &  12:37:50 &  D$\rightarrow$A &  26.294   &     23.777   &  \\
    3 &  16/2/98 &  12:26:00 &  D$\rightarrow$A &  26.450   &     24.055   &  \\
    5 &  18/2/98 &  12:46:27 &  D$\rightarrow$A &  27.164   &     24.243   &        Light snow\\
    8 &  21/2/98 &  12:05:30 &   A   &  28.004   &     24.177   &  \\
   13 &  26/2/98 &  13:46:04 &   A   &  27.104   &     23.210   &        Wind $\ge$ 5 m/s\\
   16 &   1/3/98 &  10:34:48 &   A   &  26.776   &     23.271   &  \\
   21 &   6/3/98 &  12:14:57 &   A   &  26.408   &     23.303   &  \\
   25 &  10/3/98 &  11:59:00 &   A   &  27.061   &     23.153   &  \\
   28 &  13/3/98 &  09:16:49 &   A   &  26.491   &     23.235   &  \\
   34 &  19/3/98 &  07:54:15 &   A   &  26.744   &     23.065   &  \\      
   37 &  22/3/98 &  09:42:07 &   A   &  26.783   &     22.799   &  \\      
   42 &  27/3/98 &  13:21:50 &   A   &  26.803   &     23.066   &        Wind $\ge$ 5 m/s\\
   45 &  30/3/98 &  05:41:09 &   A   &  27.136   &     22.606   &        Overcast+Light snow\\
   47 &   1/4/98 &  10:02:44 &   A   &  27.002   &     22.261   &  \\      
   50 &   4/4/98 &  15:49:50 &   A   &  26.025   &     21.456   &        Wind $\ge$ 5 m/s\\
   52 &   6/4/98 &  10:12:52 &   A   &  25.577   &     22.023   &  \\
   55 &   9/4/98 &  08:31:26 &   A   &  27.441   &     22.396   &  \\
   57 &  11/4/98 &  06:54:08 &   A   &  25.514   &     22.479   &  \\
   58 &  12/4/98 &  06:19:50 &   A   &  25.962   &     22.062   &        Wind $\ge$ 5 m/s\\
   64 &  18/4/98 &  07:26:30 &   A   &  26.111   &     22.842   &  \\      
   66 &  20/4/98 &  07:48:11 &   A   &  26.638   &     22.600   &  \\
   68 &  22/4/98 &  02:41:30 &   A   &  25.732   &     22.508   &  \\
   71 &  25/4/98 &  05:58:44 &   A   &  25.336   &     22.691   &  \\
   74 &  28/4/98 &  06:17:20 &   A   &  25.950   &     22.730   &  \\
   76 &  30/4/98 &  05:39:05 &   A   &  26.836   &     22.846   &  \\      
   77 &   1/5/98 &  00:05:54 &   A   &  27.057   &     22.481   &  \\
   77 &   1/5/98 &  09:04:22 &   A   &  26.304   &     22.637   &  \\
   80 &   4/5/98 &  07:52:35 &   A   &  25.902   &     22.224   &  \\
   84 &   8/5/98 &  05:07:09 &   A   &  26.134   &     22.354   &  \\
   88 &  12/5/98 &  06:51:32 &   A   &  26.299   &     21.904   &  \\
   93 &  17/5/98 &  10:01:50 &   A   &  25.702   &     21.712   &        Wind $\ge$ 5 m/s\\
   97 &  21/5/98 &  03:10:00 &   A   &  26.851   &     21.833   &  \\
  101 &  25/5/98 &  04:00:04 &   A   &  25.893   &     21.751   &  \\
  102 &  26/5/98 &  09:26:00 &   A   &  24.151   &     21.711   &  \\
  106 &  30/5/98 &  03:41:38 &   A   &  24.937   &     21.770   &  \\
  108 &   1/6/98 &  04:02:51 &   A   &  25.929   &     21.589   &  \\
  112 &   5/6/98 &  05:17:17 &   A   &  24.140   &     21.699   &  \\
\hline
\end{tabular}
\end{center}
\end{table*}
\addtocounter{table}{-1}

\begin{table*}
\caption{(Continued). }
\begin{center}
\begin{tabular}{crccccl}
\hline
  Epoch        & Date & Start    & Array  & $S_{8.5}^{\rm A}$ &  $S_{8.5}^{\rm B}$      & Comments\\
JD$-$2450858     && (LST)    &        &   (mJy)     &    (mJy)         &          \\ 
\hline
  116 &   9/6/98 &  04:01:50 &   A   &  23.423   &     20.862   &  \\
  118 &  11/6/98 &  03:23:33 &   A   &  24.701   &     21.342   &  \\
  121 &  14/6/98 &  06:11:18 &  BnA  &  25.760   &     22.131   &        Wind $\ge$ 5 m/s\\
  125 &  18/6/98 &  02:56:18 &  BnA  &  24.715   &     21.547   &  \\
  127 &  20/6/98 &  03:18:14 &  BnA  &  23.345   &     20.991   &        Power outage\\
  130 &  23/6/98 &  02:36:46 &  BnA  &  23.458   &     21.081   &        Wind $\ge$ 5 m/s\\
  134 &  27/6/98 &  05:20:40 &  BnA  &  23.052   &     21.005   &  \\
  140 &   3/7/98 &  03:57:00 &   B   &  23.918   &     21.100   &        Thunder+showers\\
  144 &   7/7/98 &  22:47:35 &   B   &  24.707   &     21.187   &        Thunderstorms in area\\
  146$^*$ & 9/7/98 &  02:58:38 & B   &  23.389   &     20.907   &        Showers; Erratic behavior T$_{\rm sys}$; ($\sigma_{\rm S}$=1.8\%)\\
  152 &  15/7/98 &  05:39:21 &   B   &  24.137   &     20.719   &  \\
  155 &  18/7/98 &  05:28:50 &   B   &  24.099   &     21.089   &  \\
  159 &  22/7/98 &  05:12:50 &   B   &  24.286   &     20.821   &        Distant sheet lightening\\
  162 &  25/7/98 &  05:00:56 &   B   &  24.500   &     20.700   &        Showers\\
  165$^*$  &  28/7/98 &  04:46:32 &   B   &  24.128   &     20.462   &        Wind $\ge$ 5 m/s; Showers; \\
      &          &           &       &            &             &      $\approx$50\% gradual decrease in T$_{\rm sys}$; ($\sigma_{\rm S}$=1.9\%)\\
  168 &  31/7/98 &  03:00:30 &   B   &  24.057   &     20.286   &  \\
  171 &   3/8/98 &  04:24:45 &   B   &  24.957   &     20.137   &  \\
  174 &   6/8/98 &  03:13:03 &   B   &  24.480   &     19.980   &  \\
  177 &   9/8/98 &  04:01:12 &   B   &  24.709   &     20.065   &  \\
  183 &  15/8/98 &  03:37:24 &   B   &  24.159   &     19.894   &  \\
  187 &  19/8/98 &  03:21:38 &   B   &  23.741   &     20.186   &  \\
  188$^*$ &  20/8/98 &  22:40:51 &   B   &  24.239   &     21.143   &   Wind $\ge$ 5 m/s; Showers; \\
      &          &           &       &            &             &       Erratic behavior T$_{\rm sys}$; ($\sigma_{\rm S}$=3.4\%) \\
  192 &  24/8/98 &  01:02:27 &   B   &  24.485   &     19.878   &  \\
  193 &  25/8/98 &  22:49:40 &   B   &  23.073   &     19.895   &        Wind $\ge$ 5 m/s\\
  196 &  28/8/98 &  02:22:42 &   B   &  22.522   &     19.650   &        Thunderstorms in area\\
  199 &  31/8/98 &  02:34:40 &   B   &  23.604   &     19.722   &  \\
  202 &   3/9/98 &  03:22:50 &   B   &  23.952   &     19.916   &  \\
  206 &   7/9/98 &  02:07:01 &   B   &  23.218   &     19.752   &  \\
  209 &  10/9/98 &  03:24:54 &   B   &  24.017   &     20.142   &        Wind $\ge$ 5 m/s\\
  213 &  14/9/98 &  01:39:17 &   B   &  23.235   &     20.330   &  \\
  219 &  20/9/98 &  01:16:10 &   B   &  23.369   &     20.015   &  \\
  223$^*$ &  24/9/98 &  21:57:00 &   B   &  22.766   &     20.952   &        Rapid $\approx$20\% fluctuations in T$_{\rm sys}$; ($\sigma_{\rm S}$=4.8\%) \\
  227 &  28/9/98 &  20:11:22 &   B   &  22.190   &     19.582   &  \\
  230$^*$ &  1/10/98 &  22:59:00 &   B   &  22.132   &     19.307   &        Wind $\ge$ 5 m/s; Erratic behavior T$_{\rm sys}$; ($\sigma_{\rm S}$=2.0\%)\\
  232$^*$ &  3/10/98 &  22:51:04 &   B   &  22.547   &     18.848   &        Wind $\ge$ 5 m/s; Erratic behavior T$_{\rm sys}$; ($\sigma_{\rm S}$=3.2\%)\\
  235 &  6/10/98 &  22:09:00 &   B   &  21.717   &     18.513   &  \\
  240 & 11/10/98 &  01:52:41 &   B   &  22.826   &     19.112   &        Wind $\ge$ 5 m/s\\
  243 & 14/10/98 &  01:10:54 &   B   &  22.789   &     19.080   &        Wind $\ge$ 5 m/s\\
\hline
\end{tabular}
\end{center}
\end{table*}

\subsection{External variability}

In Koopmans \& de Bruyn (1999) it is shown that most of the observed
short--term variability is of external origin (at the 14.6-$\sigma$
confidence level). A number of possible causes of this short-term
variability are examined: (i) scintillation caused by the ionized
component of the Galactic ISM and (ii) radio microlensing of a
core-jet structure by massive compact objects in the lens galaxy.
Based on Galactic scintillation models (e.g. Narayan 1992; Taylor \&
Cordes 1993; Rickett et al. 1995), one predicts a strong increase in
the modulation index towards longer wavelengths in the case of
scintillation, whereas a strong decrease is observed in the long-term
monitoring data obtained with the {\sl
Westerbork-Synthesis-Radio-Telescope} (WSRT) at 1.4 and 5 GHz
(Koopmans \& de Bruyn 1999). The quantitative decrease in the
modulation index from 5 to 1.4 GHz, seen in this WSRT monitoring data,
agrees remarkably well with that predicted from microlensing
simulations, but differs by a factor of $\approx$8 from that predicted
from the scintillation models. A strong case for the occurrence of
radio microlensing in B1600+434 can therefore be made, which can
complicate a straightforward determination of the time delay from
these light curves. However, based on two independent methods of
analysis (Sect.3.1), we are convinced that the obtained time delay is
little affected by this external variability.

\subsection{Error analysis}

The errors on the light curves are a combination of thermal noise errors and
systematic errors (e.g. modeling, self-calibration, instrumental,
atmospheric, etc.).  The noise errors on the $\ga$1\,Jy flux density
calibrators are of the order of 0.01\% after a few minutes of
integration. The noise error on each of the lens images is about
0.3\%, determined from the residual maps (i.e. the radio image after
subtracting the model of the source structure). This noise level
agrees well with the theoretically expected value for the typical
integration time of $\approx$10 min.

To estimate the systematic errors (i.e. systematic in the sense that
they affect the flux-densities of both image A and B), we compare the
two normalized light curves of B1634+627 and B1358+624. In Fig.2, we
see that both curves follow each other extremely well. Their ratio has
an rms scatter of 0.7\%, determined from fitting a Gaussian to its
distribution function.  Assuming the errors on both normalized light
curves are similar, the errors on the individual points are therefore
0.7$/\sqrt{2}$$\approx$0.5\%. This error is probably a mixture of
modeling, self-calibration and short-term atmospheric and
instrumental effects, which are hard to remove.  We conservatively
assume that the data of B1600+434 contains a similar 0.5\% error.

During the B-array observations, six points lie clearly outside the
3-$\sigma$ region (Fig.2, upper panel), whereas during the A- and
BnA-array observations, the ratio seems much more more stable. This
stability during A- and BnA-array observations is reflected in the
extremely small scatter in the measured distance between the two lens images
(Fig.4, upper panel), which is similar to the theoretical expectation
value of $\Delta r_{\rm AB}$=0.5\,mas, where we used $\Delta r_{\rm
AB} =\sqrt{2}\times \Delta\theta/(2\cdot{\rm SNR})$, with
$\Delta\theta$ being the beam size of 0.2 (0.7) arcsec in A-array
(B-array) and SNR the signal-to-noise ratio of about 1/0.003$\approx$330.

The six `outliers' are given an error equal to their difference in
normalized flux density divided by $\sqrt{2}$, which is the
expectation value if their errors are equal and drawn from a Gaussian
distribution. The errors are 1.8--4.8\%. Although this approach appears
rather ad-hoc, we will later on in the determination of a time-delay
use the light-curves both with and without these points, to
investigate the effect they have on our analysis. As it will turn out,
the effect is negligible (Sect.3.1.3).

To explain why we find these outliers, we investigated the system
temperature (T$_{\rm sys}$) as function of time.  Fast or systematic
changes in T$_{\rm sys}$ could indicate instrumental, atmospheric
(e.g. precipitation) problems or electromagnetic interference.  During
day 223 (i.e. JD$-$2450858), T$_{\rm sys}$ shows rapid changes of up to
20\% on time scales of a few minutes, which could explain the large
difference in the normalized flux density from the running mean and
the large difference in the ratio between the normalized flux
densities of B1634+627 and B1358+624 from unity. During the
observations, cumuloform type clouds were forming with $\sim$50\% sky
coverage over the array\footnote{Information obtained from the VLA
observing logs, as kept by the VLA operators.}, possibly indicating
strong interference caused by nearby thunderstorms (i.e. lightning).
Also the other five outliers, during B-array, show some erratic
behavior of T$_{\rm sys}$, although less serious than on day 223 and
typically for only several of the telescopes.  Only day 165 shows a
$\approx$50\% decrease in T$_{\rm sys}$ over a 1h time interval for
all telescopes. Because this decrease is relatively smooth,
gain-self-calibration can solve for most of the errors.  No such
behavior is found between days 45 to 52 (in A-array) for example,
which shows a much more gradual change in T$_{\rm sys}$ and maximum
differences less than 10\%. On day 45, T$_{\rm sys}$ behaves similar
to epochs with no severe data problems. Its system temperature,
however, is on average higher, which explains why the normalized flux
density is lower. The higher system temperature for each of the
telescopes is explained by the fact that it was snowing during the
observations over the entire array (100\% sky coverage). However,
because T$_{\rm sys}$ changes only gradually, the error on the
corrected flux densities of images A and B will be similar to those of
the other well behaved epochs.

Finally, we average the overlapping parts of the two normalized light
curves, such that the errors decrease by a factor $\sqrt{2}$.  Adding
all errors quadratically, we find a total error of 0.8\% on the light
curves of B1600+434 A and B in the region where the normalized flux
calibrator light curves do not overlap, and 0.7\% where they do
overlap. In Sect.3.1.4, we show that an error of 0.7--0.8\% is
statistically highly plausible, lending credibility to it.

Moreover, at a given epoch the systematic errors in the flux densities
of images A and B are the same (i.e. because of their small angular
separation of 1.4 arcsec, instrumental and atmospheric errors should
be the same, as well as initial phase calibrator errors, which have
been transferred to both images). Their flux density ratio is
therefore much better determined and has an error of only
$\sqrt{2}\times$0.3\% $\approx$0.4\%. The errors at different epochs,
however, are independent, which is important if the light curves are
shifted to determine a time delay (Sect.3.1).

\begin{figure*}[t!]
\begin{center}
  \leavevmode
\vbox{%
  \epsfxsize=\hsize
  \epsffile{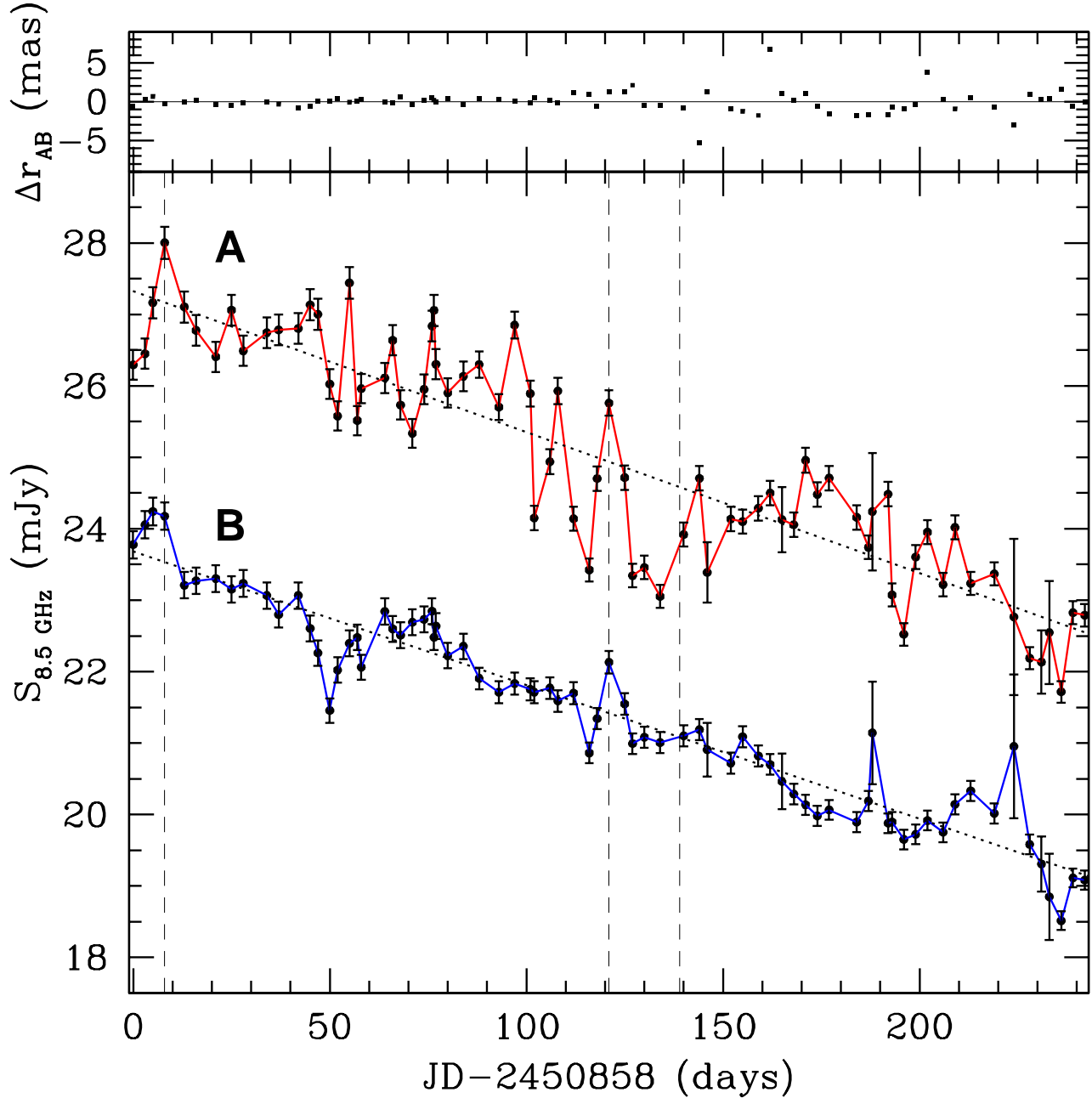}}
\end{center}
\caption{Lower panel: VLA 8.5-GHz light curves of B1600+434 A and B, starting
	Febr. 13 1998.  The error bars indicate the 1-$\sigma$ error,
	including systematic errors. The dotted lines are
	linear fits to the light curves. The vertical dashed lines indicate
	the array changes D$\rightarrow$A, A$\rightarrow$BnA and
	BnA$\rightarrow$B, respectively, from left to right. Upper panel: The
	distance between images A and B, minus its median distance of 1392\,mas
	(mean distance is also 1392\,mas).}
\end{figure*}

\begin{figure*}[t!]
\begin{center}
  \leavevmode
\hbox{%
  \epsfysize=8.5cm
  \epsffile{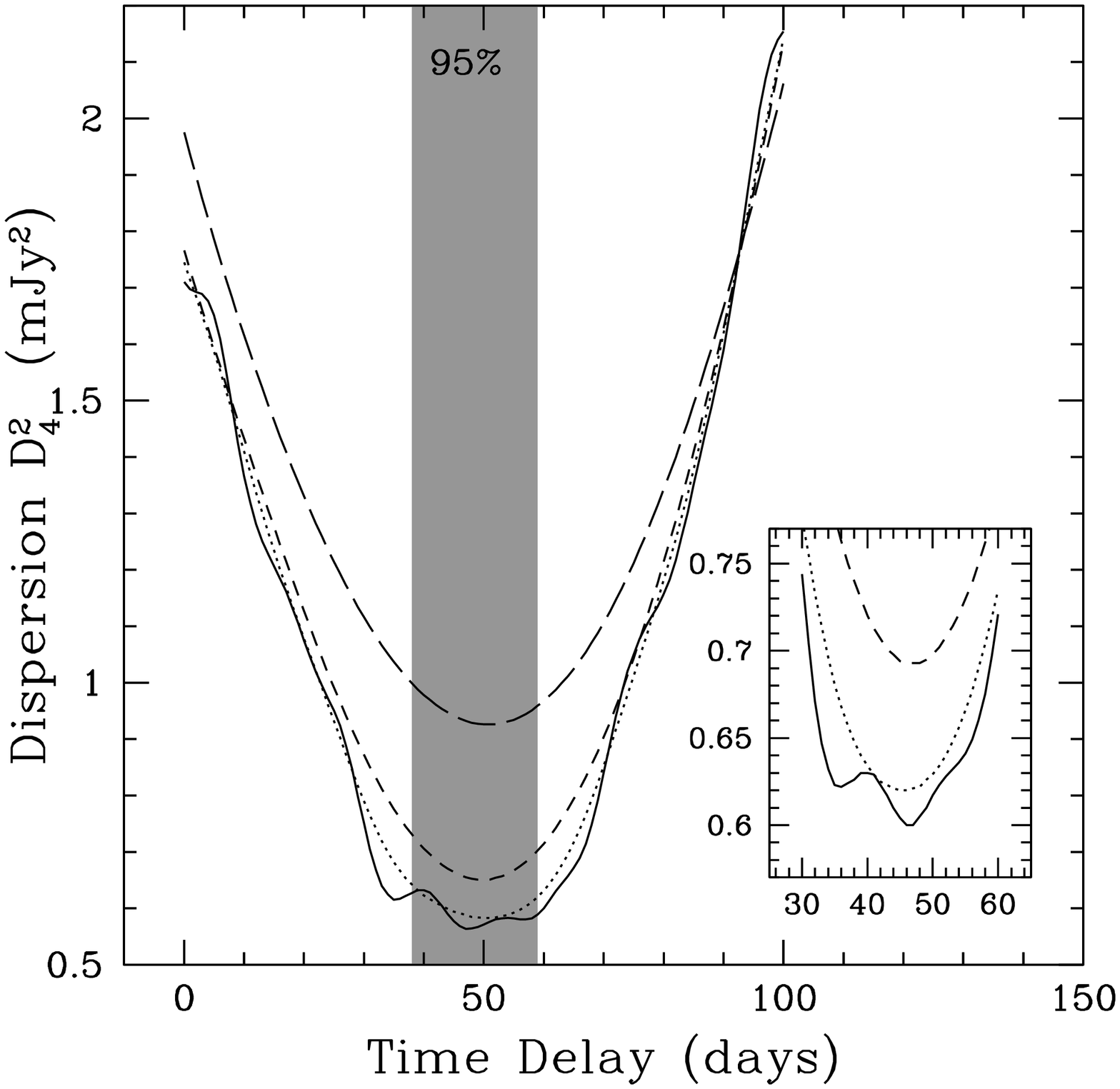}
  \epsfysize=8.5cm
  \epsffile{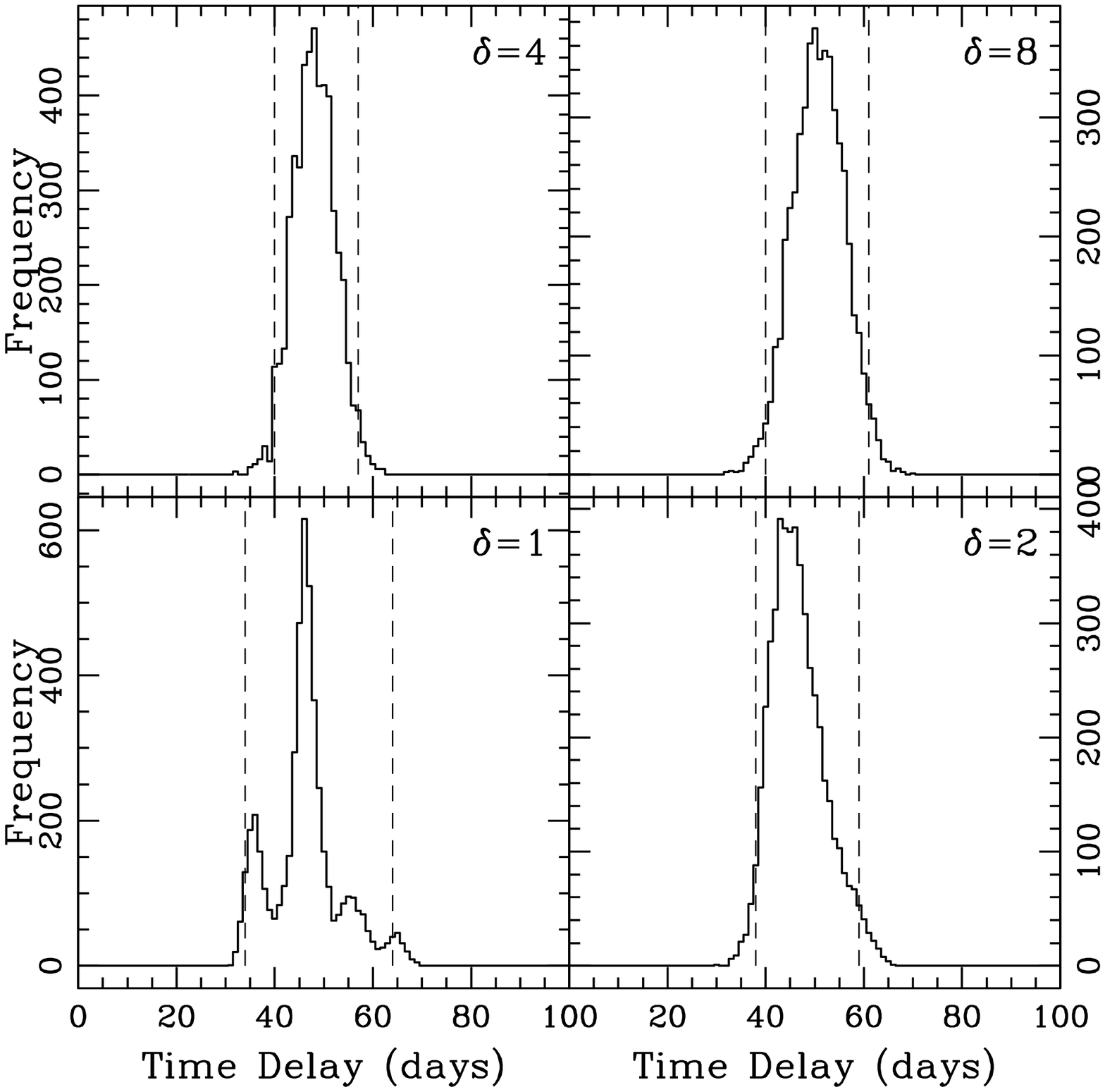}}
\end{center}
\caption{Left: Dispersion measure $D^2_{4}$ (Pelt et al. 1996) between
	the light curves of images A and B, using $r_{\rm AB}=1.212$.  The
	decorrelation time scale $\delta$=1, 2, 4 and 8 (lower to upper
	curve), in units of the average time span between observations
	(i.e. 3.3 days).  The dispersion minimizes between days 46--51. The
	shaded region indicates the 95\% statistical confidence region for the
	minimum-dispersion time delay.  Right: Time-delay PDFs from
	Monte-Carlo simulations (Sect.3.1.2). Shown are the four
	distributions for $\delta$=1, 2, 4 and 8. The bin size is 1~day and the
	dashed lines indicate the region containing 95\% of the time-delay
	PDF.}
\end{figure*}

\section{Analysis}

In this section we use the VLA 8.5-GHz light curves to put
constraints on the time delay between the two lens images.

\subsection{The time delay}

A simple estimate of the time delay could be obtained, using the
long-term gradients in both light-curves combined with the intrinsic
flux-density ratio. Fitting a straight line to both curves (Fig.4)
gives a flux-density decrease of $-$1.98$\cdot 10^{-2}$
mJy\,day$^{-1}$ for curve A and $-$1.87$\cdot 10^{-2}$ mJy\,day$^{-1}$
for curve B. Correcting the latter value by multiplying it by the
flux-density ratio of 1.212 (see below) gives $-$2.27$\cdot 10^{-2}$
mJy\,day$^{-1}$. This value is different from that of curve A by some
15\%, indicating that the rate of decrease in the flux-density
changes over the time-scale of the observations. One can therefore
not simply divide the difference in flux-density between curve A and
curve B (multiplied by the intrinsic flux-density ratio) by the rate
of decrease in flux-density to obtain a time delay.

Moreover, the rapid strong modulations seen in the light curve of
image A (Fig.4) makes interpolation questionable. This excludes the
use of either the $\chi^2$-minimization or cross-correlation methods
in determining the time delay, because the light curves have to be
resampled on a similar grid through some form of interpolation.

We have therefore chosen to use the non-parametric
minimum-dispersion method developed by Pelt et al. (1996). In
Sect.3.1.4 we will also derive the time delay using the PRH-method
from Press, Rybicki \& Hewitt (1992). As an additional constraint we
use a flux density ratio of 1.212$\pm$0.005, determined from 28 epochs
of VLA 8.5-GHz observations during a period of $\approx$4 months in
1996-1997 in which there was relatively little variability (C.B. Moore
1999, private communication). Because this period is significantly
longer than the time delay between images A and B (Sect.3.1.2), the low
rms variability implies that the above value should represent the
intrinsic flux density ratio quite closely. However, we emphasize
the preliminary nature of this value, which might still change
slightly in a final analysis.

Because the fainter image (B) lags the brighter image (A)
(Koopmans et al. 1998) and the flux densities of both images
decrease almost linearly over a period of eight months, the
flux-density ratio will on average be smaller than the flux-density
ratio of 1.212. Shifting the light curve of image B back in time will
increase the flux-density ratio between the overlapping parts of the
light curves. When the shift in time equals the time delay between the
lens images, the average flux-density ratio between the light curves
should be 1.212. Hence, if we multiply the light curve of image B
with the flux-density ratio of 1.212, we expect the minimum dispersion
between the two light curves to occur near the intrinsic time delay.

\subsubsection{Minimum dispersion method}

From the simple consideration that the observed flux-density ratio
($\approx$1.16) is smaller than the ratio of 1.212, we immediately see
that image B lags image A in time, in agreement with lens models
(Koopmans et al. 1998). Thus, the time delay $\Delta t_{\rm B-A}$ is
positive. A composite light curve is created by multiplying light
curve B with the flux-density ratio of 1.212 and shifting it backward
in time by $\Delta t_{\rm B-A}$. The dispersion of the composite light
curve is calculated as in Pelt et al. (1996), using the dispersion
measure
$$
D^2_{4}(\Delta t)= \frac{\sum_{n=1}^{N-1}\sum_{m=n+1}^{N} S_{\rm m,n}W_{m,n}G_{m,n} (C_{\rm
	n} - C_{\rm m})^2} {\sum_{m=n+1}^{N} S_{\rm m,n}W_{m,n}G_{m,n}}
$$
where $C_{\rm n}$ is the $n$-th point on the composite light curve
and $G_{m,n}$=1 (0), if $C_{\rm n}$ and $C_{\rm m}$ are from 
different (the same) light curves. 
We calculate the dispersion for all time delays $\Delta
t_{\rm B-A}$=0--100 days, in steps of 1 day.  This process is performed
for four different decorrelation time scale, $\delta$=1,2,4,8, in
units of the average time span between epochs, i.e. 3.3~days. We use
a decorrelation weight function $S_{\rm m,n}=\exp[-(t_{\rm m}-t_{\rm
n})^2/(2\Delta^2)]$, where $\Delta=3.3\times\delta$ is the
decorrelation time scale in days. The statistical weights are
$W_{m,n}=(W_{n}W_{m})/(W_{n}+W_{m})$, where  $W_{i}=1/\sigma_{i}^2$
and $\sigma_i$ the 1-$\sigma$ error on the flux density at the $i$-th
epoch. 

In Fig.5 the dispersion $D^2_{4}$ is plotted versus the time delay,
for $\delta$=1,2,4 and 8. The dispersion minimizes near a time delay
of 46 to 51 days.

\subsubsection{Median time delay and statistical error range}

To determine a statistical confidence region for the time delay, we
performed Monte-Carlo simulations for $\delta$=1,2,4 and 8. First, we
re-sampled the light curves, using the sampling-interval distribution
determined from the VLA observations. Because the light curves exhibit
variability due to external causes (Sect.2.6), we do not create a
composite light curve, by combining the image light curves, as has
been done for B0218+357 (Biggs et al. 1999) and B1608+656 (Fassnacht
et al. 1999).  Such a light curve only resembles the true underlying
light curve, if all variability were intrinsic to the source, which is
not the case for B1600+434. We therefore linearly interpolate the
observed light curves and errors to obtain the flux densities and
errors at the re-sampled intervals. Subsequently, Gaussian distributed
errors are added to (i) each point on the re-sampled light curves
(1-$\sigma$ equal to the interpolated flux-density error) and (ii)
the assumed intrinsic flux-density ratio ($\sigma_r$=0.005). The
process described above was repeated 5000 times for each decorrelation
time scales. The delays were stored, where $D^2_{4}$ minimizes.  The
resulting time-delay probability distribution functions (PDF) for the
four decorrelation time scales are shown in Fig.5.
From the PDFs of $\Delta t_{\rm B-A}$, we determine the median values
for the time delay and the statistical confidence regions. The final
result of this procedure, for the different decorrelation scales, are
listed in Table~3.

\begin{table}
\caption{The median time delays ($\Delta t_{\rm B-A}$) between
	B1600+434 A and B, determined from the PDFs for different
	decorrelation time scales ($\delta$). The 68\% and 95\%
	statistical confidence regions are listed in the last two columns,
	respectively.}
\begin{center}
\begin{tabular}{cccc}
\hline
$\delta$  & $\Delta t_{\rm B-A}$ (d) & 68\% (d) & 95\% (d)\\
\hline
1 & 46 & 38-53 & 34-64\\
2 & 46 & 41-52 & 38-59\\
4 & 48 & 44-52 & 40-57\\
8 & 51 & 45-56 & 40-64\\
\hline
\end{tabular}
\end{center}
\end{table}

For $\delta$=1, multiple strong peaks are found (Fig.5).  For
$\delta$$>$1, only one peak is found. The small decorrelation time
scale for $\delta=1$ makes the dispersion measure especially sensitive
to the modulations in both light curves. These peaks, however, are
much stronger in the light curve of image A and the result of
external causes (Sect.2.6; Koopmans \& de Bruyn, 1999). Hence, a
somewhat larger decorrelation time scale will give a better estimate
of the time delay, because it is less sensitive to these modulations
(i.e. it averages over these modulations). For very large
decorrelation time scales, however, one becomes sensitive to the fact
that both light curves show a long-term gradient. The gradient
introduces a systematic difference in the flux-density level between
points on the two different light curves, if they are separated by a
large time interval. This artificially increases the dispersion with
increasing $\delta$, as can be seen in Fig.5. This effect also seems
to increase the width of the 95\% statistical confidence interval. The
intermediate decorrelation time scales ($\delta$=2 and 4) therefore
seem to give a better estimate of the time delay, as it avoids most of
these problems.

For the median value of the time delay we take the average of
$\delta$=2 and $\delta$=4, which seem to give the most stable solution (see
also Sect.3.1.3 and Fig.6). For the 68\% and 95\% statistical
confidence regions, we conservatively take their combined maximum
ranges. Hence,

\begin{eqnarray*}
\begin{array}{lllll}
	\Delta t_{\rm B-A} & = & 47^{+5}_{-6} & {\rm d} & \mbox{~~~(68\%)}\\ 
			   & = & 47^{+12}_{-9}& {\rm d} & \mbox{~~~(95\%)},
\end{array}
\end{eqnarray*}

\noindent which we take as our best estimate of the time delay between the 
images in the gravitational lens system B1600+434 and the statistical 
confidence intervals.

\subsubsection{Systematic uncertainties in the time delay}

Several systematic uncertainties remain, which we will investigate below: 

\begin{enumerate}

\item{The first has to do with the six outliers on the
      light-curves (Sect.2.7). Although we gave them significantly
      larger errorbars, they can still affect the determination of the
      time delay.}

\item{The choice of the decorrelation time scale seems to influence
      the median time delay (Sect.3.1.2). Larger decorrelation time
      scales give larger median time delays.}

\item{The flux density ratio that we used in our analysis (Sect.3.1) is
      preliminary and the error was assumed to be Gaussian,
      which might not be the case.}

\end{enumerate}
To address the first two points, we ran Monte-Carlo simulations (500
redistributions), for decorrelation time scales of $\Delta$=3,5,...,25
d, using {\sl all} epochs shown in Fig.4. We repeated this without the
six outliers (Sect.2.7). The results are shown in Fig.6. The
values for the median time delays range between 43 to 51 d, for the
assumed range of decorrelation time-scales. The width of the 95\%
statistical confidence interval seems to minimize in the range
$\Delta$$\approx$10--15 d, as was already noted in the previous
section.

To estimate the effect of a wrongly chosen intrinsic flux-density
ratio, we also ran models for $r_{\rm AB}$=1.202 and 1.222, which are the
assumed intrinsic flux-density ratio plus-minus twice its estimated
error. The first value decreases the median time delay systematically
by about 8 days, whereas the latter value increases it by about 7
days. We thus take the range of about $-$8 to +7 days as a good
indication of the maximum systematic error range.

\begin{figure}[t!]
\begin{center}
  \leavevmode
\vbox{%
  \epsfxsize=\hsize
  \epsffile{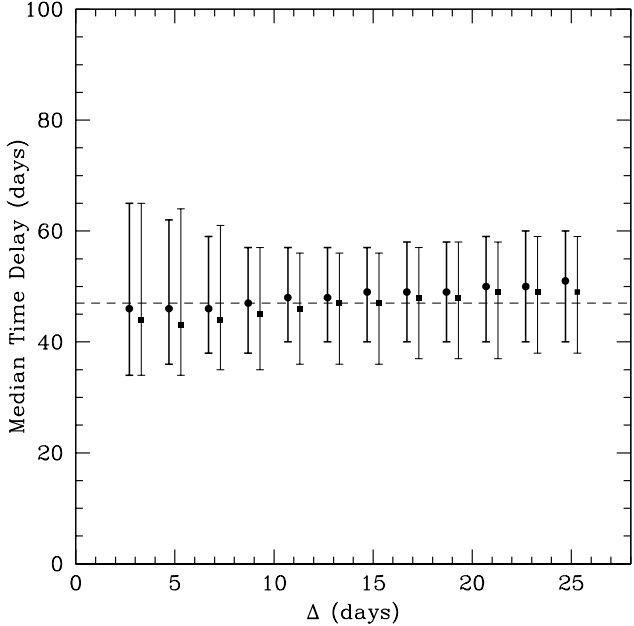}}
\end{center}
\caption{The median time delay determined from the time delay PDFs,
as function of the decorrelation time scale $\Delta$. The
errorbars indicate the region containing 95\% of the PDF. The circles
indicate the time delays, using {\sl all} epochs of the light curves
shown in Fig.4. The squares indicate the time delays, omitting the
six strongest ``outliers'' (Sect.3.1.3). The dashed line indicates 
a time delay of 47 days (Sect.3.1.2).}
\end{figure}

\subsubsection{The PRH method}

We also applied the PRH method (Press, Rybicki \& Hewitt 1992), to
address the question if the time delay obtained in Sect.3.1.2 might
critically depend on the method that was used.  We use the
implementation of this method in the {\sf ESO-MIDAS} (version 96NOV)
data-reduction package.  We tried several different analytical
functions to describe the structure functions of the observed light
curves and find no strong dependence of the final result
(i.e. differences of $\la$1 day in the time delay) on the precise
functional form of the structure function.  We exclude the six
outliers (Sect.2.7) from the analysis and find a delay of $\Delta
t_{\rm B-A}$=$48^{+2}_{-4}$ d (2-$\sigma$), where the formal error is
derived by varying the time-delay until $\Delta\chi^2$ has increased
by 4.0. This error is smaller than that derived in Sect.3.1.2, because
it does not include all the uncertainties that we took into account in
the Monte-Carlo simulations. However, we see that both the
minimum-dispersion method and the PRH method give consistent results
within their respective 1-$\sigma$ error regions.  Thence, we
conclude that the resulting value of $\Delta t_{\rm B-A}$ does not
strongly depend on the method that is used, at least in the case of
B1600+434.

We furthermore find a minimum-$\chi^2$ value of $\approx$134 at $\Delta
t_{\rm B-A}$=48 d for 137 degrees of freedom, which is statistically
highly plausible. This lends strong credibility to our error
analysis (Sect.2.7), suggesting that the error on the individual
flux-density points are indeed 0.7--0.8\% for most epochs.

\subsection{The Hubble parameter and slope of the radial mass profile
of the lens-galaxy dark-matter halo}

To estimate a tentative value for the Hubble parameter (H$_0$), we use
the mass models for B1600+434 from Koopmans et al. (1998). In doing
this, one should keep in mind the difficulties and degeneracies that
were mentioned in Sect.1. The values derived here should therefore be
regarded as indicative, as long as no better constraints on the radial
mass profile of the lens galaxy are obtained.

Spectroscopic observations of the lens system on 1998 April 20 with
the W.M. Keck-II telescope, have recently confirmed the assumption
that the nearby companion galaxy (G2) has the same redshift ($z$=0.41;
Fassnacht et al., in preparation) as the lensing galaxy (G1). For a
lens galaxy with an oblate isothermal dark-matter halo, the relation
between the time delay and Hubble parameter was then found to be $
\mbox{H}_0=50\times [(54^{+11}_{-9}\mbox{ d})/ {\Delta t_{\rm B-A}}] $
km~s$^{-1}$~Mpc$^{-1}$, for $\Omega_{\rm m}$=1 and
$\Omega_{\Lambda}$=0. The errors indicate the maximum range of the
isothermal-model time delays from Koopmans et al. (1998). Recently,
the time-delay dependence of H$_0$ found for this isothermal mass
model was corroborated by Maller et al. (1999), who did a similar
analysis of B1600+434, using a deep NICMOS-F160W {\sl HST} exposure.
Combining this relation with the median time delay (Sect.3.1.2), the
Hubble parameter then becomes $\mbox{H}_0=57^{+14}_{-11}$
km~s$^{-1}$~Mpc$^{-1}$ (95\%), for $\Omega_{\rm m}$=1 and
$\Omega_{\Lambda}$=0.  A maximum systematic error between $-$15 to +26
km s$^{-1}$ Mpc$^{-1}$ is estimated from the combination of model and
systematic time-delay errors. This error does not include the
uncertainty in the slope of the radial mass profile.  In fact, for a
Modified Hubble Profile (MHP) halo mass model we find a significantly
higher value for the Hubble parameter, H$_0$=74$^{+18}_{-15}$ km
s$^{-1}$ Mpc$^{-1}$ (95\%), with a maximum systematic error between
$-$22 to +22 km s$^{-1}$ Mpc$^{-1}$. For a flat universe with
$\Omega_0$=0.3 and $\Omega_{\Lambda}$=0.7, these values of H$_0$
increase by 5.4\%. At this points B1600+434 is not a GL system from
which H$_0$ can be constrained reliably.

In Wucknitz \& Refsdal (1999) it was shown that in general the time
delay is a strong function of the slope of the radial mass profile of
the mass distribution of the lens galaxy. Because the lens-image
properties can be recovered to first order for a range of different
slopes, the expected time delay from a GL system as function of H$_0$
is also a strong function of this slope. This degeneracy between the
time delay and the slope makes an accurate determination of H$_0$ from
GL systems difficult, especially in two-image lens systems like
B1600+434. Hence, if H$_0$ can be determined with greater accuracy
($\la$10\% error), the measured time delay can be used to constrain
the slope of the radial mass profile of the dark-matter halo.

\section{Conclusions}

We have monitored the CLASS gravitational lens B1600+ 434 at 8.5 GHz
with the VLA in A and B-arrays, during the period from February to
October 1998.  The light curves show a nearly linear decrease of
18-19\% in the flux density of both lens images over this
period. However, image A also shows rapid variability (up to 11\%
peak-to-peak) on scales of days to weeks, whereas image B shows
significantly less short-term variability (upto 6\%
peak-to-peak). The short-term variability occurs over an observing
period that is much longer than any conceivable time delay.

In Koopmans \& de Bruyn (1999) it is shown that the short-term
variability is predominantly of external origin.  Two plausible
explanations of this external variability are suggested: scintillation
caused by the ionized component of the Galactic ISM or radio
microlensing of a core-jet structure by massive compact objects in the
lens galaxy. Both possibilities are examined in more detail in
Koopmans \& de Bruyn (1999). A comparison of the result from the VLA
8.5-GHz, presented in this paper, and multi-frequency (1.4 and
5-GHz) WSRT monitoring data with the expected dependence of
scintillation on frequency (e.g. Narayan 1992; Taylor \& Cordes 1993;
Rickett et al. 1995), shows that the scintillation hypothesis
underestimates the short-term rms variability at 1.4 GHz by a factor
$\approx$8.  Within the uncertainties, the microlensing hypothesis
predicts the correct frequency-dependence of the short-term rms
variability as function of frequency. The radio-microlensing
hypothesis therefore seems most viable at present.

From the VLA 8.5-GHz light curves, we determined a median time delay
of $\Delta{\rm t}_{\rm B-A}$=47$^{+12}_{-9}$ days (95\% statistical
confidence) between the lens images. A maximum systematic error
between $-$8 and +7 d is estimated. We used the minimum-dispersion
method from Pelt et al. (1996), but find the same time-delay from the
PRH-method from Press et al. (1992).

Combining this with the isothermal lens mass models from Koopmans et
al. (1998), the Hubble parameter would become H$_0$=57$^{+14}_{-11}$
km s$^{-1}$Mpc$^{-1}$ (95\%) for $\Omega_{\rm m}$=1 and
$\Omega_{\Lambda}$=0. A maximum systematic error between $-$15 and +26
km s$^{-1}$ Mpc$^{-1}$ is estimated.  Similarily, the MHP mass models
would give H$_0$=74$^{+18}_{-15}$ km s$^{-1}$ Mpc$^{-1}$ (95\%), with
a maximum systematic error between $-$22 and +22 km s$^{-1}$
Mpc$^{-1}$, for the same cosmological model. We hope to improve on the
determination of this time delay with an ongoing three-frequency VLA
monitoring campaign (June 1999 to Feb. 2000). Because of the
degeneracy between the slope of the radial mass profile and the
expected time delay between the lens images as function of H$_0$, the
above-given values of H$_0$ should be regarded as indicative.

If H$_0$ can be determined accurately from independent methods and no
extra constraints on the lens model can be found, it is more
interesting to use that observed time delay to constrain the slope of
radial mass profile of the dark-matter halo around the lensing 
edge-on spiral galaxy in B1600+434.

\begin{acknowledgements}

We like to thank Chris Moore for useful
discussions and several good suggestions to improve the manuscript.
We thank Phillip Helbig, Peter Wilkinson and Ian Browne for carefully
reading the manuscript and giving suggestions for improvement. We
thank Jaan Pelt and Mark Neeser for helping to implement the PRH method in
{\sf MIDAS}. LVEK and AGdeB acknowledge the support from an NWO
program subsidy (grant number 781-76-101). This research was supported
in part by the European Commission, TMR Program, Research Network
Contract ERBFMRXCT96-0034 `CERES'. The National Radio Astronomy
Observatory is a facility of the National Science Foundation operated
under cooperative agreement by Associated Universities, Inc.  The
Westerbork Synthesis Radio Telescope (WSRT) is operated by the
Netherlands Foundation for Research in Astronomy (ASTRON) with the
financial support from the Netherlands Organization for Scientific
Research (NWO). This research has made use of data from the University
of Michigan Radio Astronomy Observatory which is supported by funds
from the University of Michigan.

\end{acknowledgements}

\end{document}